\documentclass[journal]{IEEEtran}

\usepackage{xcolor,soul,framed} 

\colorlet{shadecolor}{yellow}
\usepackage{color,soul}
\usepackage[pdftex]{graphicx}
\graphicspath{{../pdf/}{../jpeg/}}
\DeclareGraphicsExtensions{.pdf,.jpeg,.png}

\usepackage[cmex10]{amsmath}
\usepackage{array}
\usepackage{mdwmath}
\usepackage{amssymb}
\usepackage{mdwtab}
\usepackage{eqparbox}
\usepackage{url}
\usepackage{multirow}
\usepackage{tabularx}
\usepackage{threeparttable}
\usepackage{booktabs}

\begin{document}
\bstctlcite{IEEEexample:BSTcontrol}
    \title{Multi-Contrast Super-Resolution MRI Through a Progressive Network}
  \author{Qing~Lyu,
      Hongming~Shan,
      Ge~Wang$^\ast$,~\IEEEmembership{Fellow,~IEEE}
  \thanks{Asterisk indicates corresponding author.}
  \thanks{Q. Lyu, H. Shan, and G. Wang are  with Department
of Biomedical Engineering, Rensselaer Polytechnic Institute, Troy, NY, 12180 USA (e-mail: \{lyuq, shanh, wangg6\}@rpi.edu)}
  }  


\maketitle

\begin{abstract}
{M}agnetic resonance imaging (MRI) is widely used for screening, diagnosis, image-guided therapy, and scientific research. A significant advantage of MRI over other imaging modalities such as computed tomography (CT) and nuclear imaging is that it clearly shows soft tissues in multi-contrasts. Compared with other medical image super-resolution (SR) methods that are in a single contrast, multi-contrast super-resolution studies can synergize multiple contrast images to achieve better super-resolution results. In this paper, we propose a one-level non-progressive neural network for low up-sampling multi-contrast super-resolution and a two-level progressive network for high up-sampling multi-contrast super-resolution. Multi-contrast information is combined in high-level feature space. Our experimental results demonstrate that the proposed networks can produce MRI super-resolution images with good image quality and outperform other multi-contrast super-resolution methods in terms of structural similarity and peak signal-to-noise ratio. Also, the progressive network produces a better SR image quality than the non-progressive network, even if the original low-resolution images were highly down-sampled.
\end{abstract}

\begin{IEEEkeywords}
Magnetic resonance imaging (MRI), super-resolution (SR), multi-contrast, neural network, progressive neural network
\end{IEEEkeywords}

%
\IEEEpeerreviewmaketitle

\section{Introduction}

\IEEEPARstart{M}{agnetic} resonance imaging (MRI) is one of the most widely used medical imaging modalities. Compared with other modalities such as computed tomography (CT) and nuclear imaging, MRI is advantageous in providing clear tissue structure and functional information without inducing ionizing radiation. With pulse sequences, the MRI system can be flexibly configured to generate multi-contrast images like T1, T2, and proton density (PD) weighted images, which contain important physiological and pathological features. However, a major shortcoming of current MRI system is that it is difficult to obtain high-resolution (HR) MR images in clinical applications due to the trade-off among the system cost-effectiveness and signal-to-noise ratio \cite{plenge2012super,van2012super}. Clinically, in order to obtain HR MR images, patients are required to remain stable in the gantry for long time, which intensifies patients’ discomfort and inevitably introduces motion artifacts that compromise image quality.

Super-resolution (SR) techniques have a great utility in improving MR image quality without any hardware modification. Current SR approaches can be categorized into single-image super-resolution (SISR) methods and multi-contrast super-resolution (MCSR) methods. The SISR approach was extensively studied in the past a few decades, which aims to improve low-resolution (LR) images obtained in a single contrast mode. Currently, there are many SISR methods in the literature. From model-based methods like interpolation algorithms \cite{park2003super} and iterative deblurring algorithms \cite{hardie2007fast,manjon2010non} to learning-based methods such as dictionary learning methods \cite{yang2010image,zeyde2010single,timofte2014a+}. Many impressive SR results were achieved but they are still insufficiently mature for clinical use. In recent years, SISR becomes a hot topic in the deep learning field. A number of neural network-based SR models were proposed. According to different model designs, these models can be categorized into linear models such as SRCNN  \cite{dong2015image} and VDSR \cite{kim2016accurate}, residual models like CARN \cite{ahn2018fast} and REDNet \cite{mao2016image}, recursive models like DRCN \cite{kim2016deeply} and DRRN \cite{tai2017image}, densely connected models RDN \cite{zhang2018residual} and D-DBPN \cite{haris2018deep}, attention-based models with SelNet \cite{choi2017deep} and RCAN \cite{zhang2018image} as examples, progressive models SCN \cite{wang2015deep} and LapSRN \cite{lai2017deep}, and generative adversarial network (GAN) models EnhanceNet \cite{sajjadi2017enhancenet} and SRGAN \cite{ledig2017photo}. 

Clinically, T1, T2 and PD weighted images are often generated together for diagnosis with complementary information. Although weighted images are only good at showing a certain type of tissues, they reflect the same anatomy, and can provide synergy when used in combination. Based on this consideration, the MRI MCSR approach is naturally conceivable to use MRI multi-contrast images collectively for SR imaging superior to SISR results. Typically, MCSR imaging may achieve super-resolution results for a specific contrast image aided by HR images with other contrasts. Relevant information in those reference images is extracted to guide recovery of details in the image of interest. Similar to the SISR methods, current MCSR methods can also be categorized into model-based methods and learning-based methods. Examples of model-based methods are based on non-local mean \cite{jafari2014mri}, total variation \cite{brudfors2018mri}, edge gradient \cite{zheng2018multi}, shareable information \cite{gong2015promise}, and similarity \cite{manjon2010mri,zheng2017multi}. The methods using dictionary learning \cite{lu2015mr} and convolutional neural network \cite{zeng2018simultaneous} are representatives of the learning-based methods.

In this paper, we propose a one-level non-progressive network for low up-sampling factor MCSR and a two-level progressive neural network for high up-sampling factor MCSR. The main contributions of this paper are: 1) the two-level progressive neural network based on the Wasserstein generative adversarial network with gradient penalty (WGAN-GP) architecture that can achieve excellent MCSR results in the case of a high up-sampling factor; 2) the finding that among multiple multi-contrast information integration strategies, information combined in a high-level feature space gives the best MCSR results; and 3) the composite loss involving mean-squared-error (MSE), perceptual loss, and texture matching loss to ensure that generated images can recover texture details and faithful to the ground truth.

\begin{figure*}[!ht]
\centering
\includegraphics[width=7.16in]{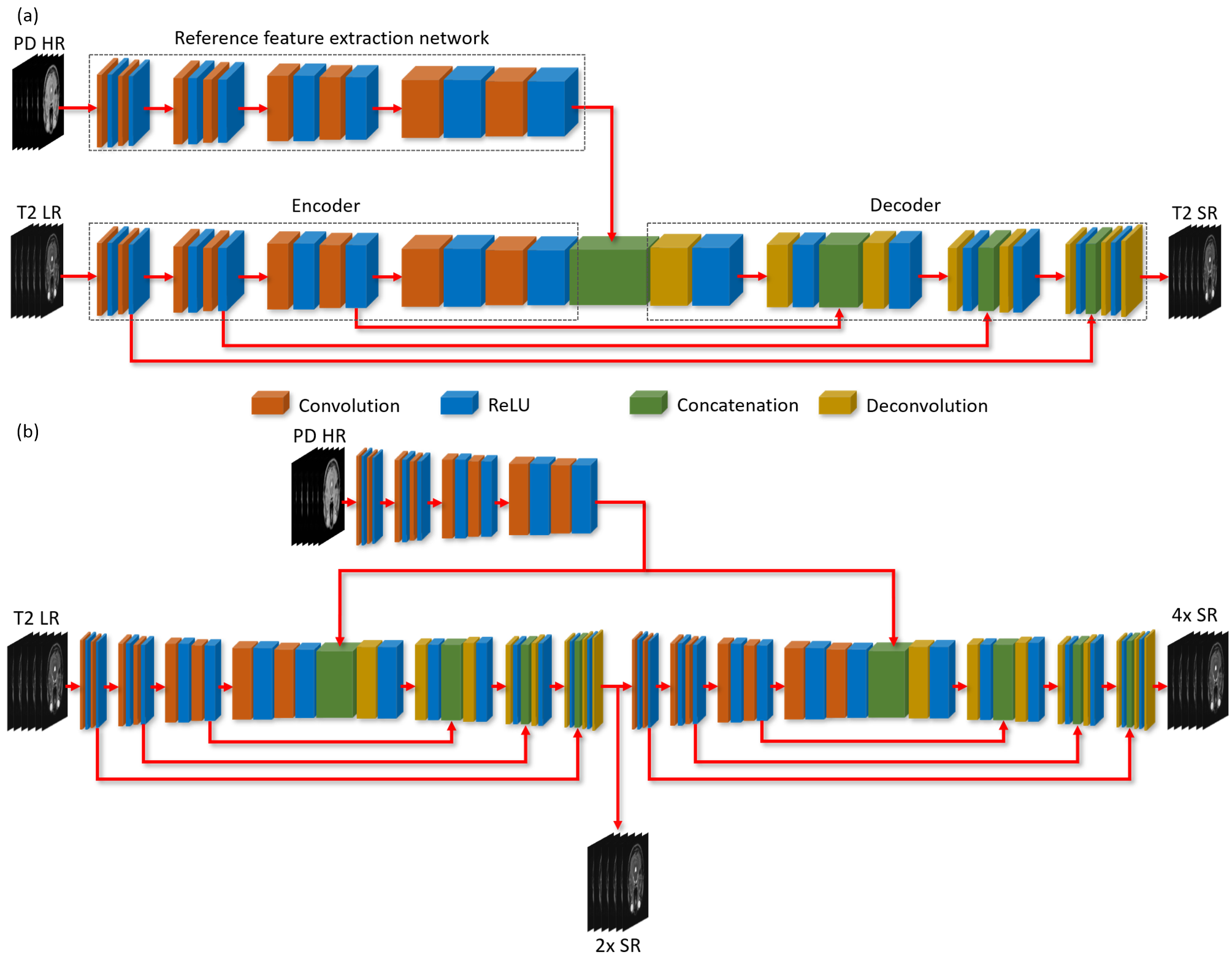}
\caption{Two proposed network architectures for MCSR. (a) The generator structure of the proposed one-level non-progressive model, which contains an encoder-decoder network and a reference feature extraction network. (b) The generator structure of the proposed two-level progressive model.}
\label{fig_1}
\end{figure*}

\section{Methodology}

\subsection {Overall SR Process}
Given an HR MR image $I_{HR} \in \mathbb{R}^{n \times n}$ of the size $n \times n$, the LR image $I_{LR} \in \mathbb{R}^{n/s \times n/s}$ obtained after the down-sampling process with a factor $s>1$ will be of the size $n/s \times n/s$. This down-sampling process $f$ can be expressed as:
\begin{equation}\label{equ_1}
    I_{LR} = f(I_{HR}) = \varphi(I_{HR}) + \epsilon,
\end{equation}
where $\varphi$ denotes the down-sampling or blurring function, and $\epsilon$ represents the system noise. Theoretically, an SR process is to find the inverse solution $f^{-1}$ of the original down-sampling function $f$. As SR imaging is an ill-posed problem, it is impossible to find the exact inverse solution, and only approximate solutions can be obtained. The goal of the SR imaging process is to find a most desirable inverse function $g$ of the ideal inverse solution $f^{-1}$. Then, the SR process can be expressed as:
\begin{equation}\label{equ_2}
    I_{SR} = g(I_{LR}) \approx I_{HR}.
\end{equation}

To obtain such an approximate solution $g$, image priors should be used. In our previous SISR study \cite{lyu2019mri}, we proposed an ensemble-learning-based deep learning framework to achieve MRI SR imaging with complementary image priors. Although a dataset with rich image features was built for single-contrast images, the amount of prior information gathered from single-contrast images is still limited. Therefore, in this study we take advantage of multi-contrast images for MRI SR imaging, which should contain more prior information than single-contrast images. Specifically, we propose a deep learning-based MCSR method for SR T2 weighted imaging by incorporating HR PD or T1 weighted images as reference images.

\subsection {Down-sampling and Zero-filling}
In our previous study \cite{lyu2019mri}, LR MRI images were obtained by down-sampling T2-weighted images in the frequency domain. We first converted the original T2 image of size $256 \times 256$ into the \textit{k}-space. Then, we cropped data and only kept data points in a central low frequency region. For the down-sampling factors 2, 3 and 4, the central 25\%, 11.1\% and 6.25\% data points were kept. All the peripheral data points were zeroed out, which is called zero-filling. Then, we used the inverse Fourier transform to convert the modified data back into the image domain to produce LR images. Through such a down-sampling and zero-filling process, we degraded the image quality and kept the image size unchanged. 

\subsection {One-Level Non-progressive Network}
The proposed one-level non-progressive network is based on the Wasserstein generative adversarial network with gradient penalty (WGAN-GP) \cite{gulrajani2017improved} which includes a generator and a discriminator. As shown in Fig. \ref{fig_1} (a), the generator consists of an encoder-decoder network and a reference feature extraction network. 

The encoder-decoder network \cite{shan20183} mainly contributes to the improvement of image quality. The encoder has 8 sequential convolutional layers, each of which is followed by a rectified linear unit (ReLU). The number of these convolutional filters are 32, 32, 64, 64, 128, 128, 256, and 256, respectively. Each convolutional layer uses $3 \times 3$ convolution kernels with a stride of 1 and a padding of 0. The encoder condenses information and extracts features layer by layer. The decoder consists of 8 sequential transposed convolutional layers, each of which is also followed by a ReLU. The numbers of filters in each transposed convolutional layer are 256, 128, 128, 64, 64, 32, 32, 1, respectively. Differing from convolutional layers, transposed convolutional layers enlarge feature maps through transposed convolution filters of kernel size $3 \times 3$ with a stride of 1 and a padding of 0. The decoder gradually dilates information layer by layer. Between the encoder and the decoder, there are three skip connections. Feature maps extracted in the encoder are transferred to the decoder through the skip connections. These skip connections can boost the training process through utilizing the similarity between input and output images. The discriminator includes three blocks followed by two fully-connected layers. Each block contains two convolutional layers, two ReLU activation functions, and a $2 \times 2$ max-pooling layer. Each convolutional layer contains $3 \times 3$ convolution kernels with a stride of 1 and a padding of 1. The numbers of convolutional filters are 64, 64, 128, 128, 256, and 256, respectively. The numbers of neurons in the two fully-connection layers are 128 and 1 respectively.

The reference feature extraction network is utilized to extract feature maps from the reference images. These extracted features are then fed into the encoder-decoder network. The structure of the reference feature extraction network is the same as the encoder in the encoder-decoder network. 

\subsection {Objective Function}
The objective function of the generator contains four parts: the adversarial loss $\mathcal{L}_{adv}$, MSE loss $\mathcal{L}_{mse}$, perceptual loss $\mathcal{L}_{per}$, and texture matching loss $\mathcal{L}_{txt}$.

The adversarial loss is used to train the generator to produce indistinguishable results from original HR images so that it can fool the discriminator. For a WGAN-GP model, the adversarial loss can be expressed as:
\begin{equation}\label{equ_3}
    \mathcal{L}_{adv} = -\mathbb{E}_{x\sim I_{LR}}[D(G(x))],
\end{equation}
where $G$ represents the generator, $D$ stands for the discriminator, and $I_{LR}$ is the input LR image of the generator.

The MSE is one of the most widely used fidelity terms in SR studies. It evaluates the similarity between generator outputs and ground truths in the image space. MSE can greatly improve the signal-to-noise ratio of generated results \cite{wang2004image}. MSE loss can be expressed as
\begin{equation}\label{equ_4}
    \mathcal{L}_{mse} = \mathbb{E}_{x\sim I_{LR},y\sim I_{HR}}{\Vert G(x) - y\Vert}^2.
\end{equation}

Although MSE can increase generator outputs with high PSNR and structural similarity to ground truth, it tends to produce over-smoothing SR results \cite{dosovitskiy2016generating,johnson2016perceptual}. To overcome this problem, the perceptual loss is included in the objective function. Different from the MSE loss that evaluates image similarity at the pixel-level, the perceptual loss measures image similarity in a high-level feature space. We used the pre-trained VGG16 model \cite{simonyan2014very} to extract feature maps. Specifically, feature maps from the $2^{nd}$, $4^{th}$, $7^{th}$, and $10^{th}$ convolutional layers were selected for calculation of the perceptual loss. All feature maps equally contribute to the perceptual loss. The perceptual loss can be expressed as the MSE of feature maps between the generator outputs and the ground truth:
\begin{equation}\label{equ_5}
    \mathcal{L}_{per} = \mathbb{E}_{x\sim I_{LR},y\sim I_{HR}}{\Vert \phi(G(x)) - \phi(y)\Vert}^2,
\end{equation}
where $\phi$ represents extracted feature maps.

Proposed by Gatys et al. \cite{gatys2016image}, the texture matching loss contributes to the texture similarity between the generator output and ground truth \cite{sajjadi2017enhancenet}. The texture matching loss calculates the Gram matrix of the generator result and the ground truth in a feature space. The Gram matrix measures feature correlations of multiple layers. Texture information can be kept by decreasing the Gram matrix difference between the generator result and the ground truth. This texture matching loss is also called the style loss in image style transfer studies because the Gram matrix can represent the style of an image. This type of characteristics is suitable for MCSR studies as well. In MCSR images, there exists a style transfer between different contrast images. This loss function ensures correct features are used in this style transfer process. The texture matching loss is the MSE of the Gram matrix between the generator output and the ground truth in a feature space:
\begin{equation}\label{equ_6}
    \mathcal{L}_{txt} = \mathbb{E}_{x\sim I_{LR},y\sim I_{HR}}{\Vert GM(G(x)) - GM(y)\Vert}^2,
\end{equation}
where $GM(F) = FF^T \in \mathbb{R}^{nl \times ml}$ stands for the Gram matrix defined as the outer product of feature map matrix $F \in \mathbb{R}^{nl \times ml}$ and its transpose matrix $F^T$. $nl$ is the number of feature maps in the convolution layer, and $ml$ stands for the height times the width of the feature map. In this study, feature maps involved in the texture matching loss are the same as those used for the perceptual loss.

In total, the objective function of the generator is:
\begin{equation}\label{equ_7}
    \min_{\theta_G}\mathcal{L}_G = \mathcal{L}_{adv} + \lambda_1\mathcal{L}_{mse} + \lambda_2\mathcal{L}_{per} + \lambda_3\mathcal{L}_{txt},
\end{equation}
where $\theta_G$ presents the trainable parameters of the generator.

The objective function of the discriminator is the same as the objective function of the original WGAN-GP model, which is defined as follows:
\begin{equation}\label{equ_8}
\begin{split}
    \min_{\theta_D}\mathcal{L}_D = & \mathbb{E}_{x\sim I_{LR}}[D(G(x))] - \mathbb{E}_{\hat{x}\sim I_{HR}}[D(\hat{x})] \\
    & + \lambda_4\mathbb{E}_{\Tilde{x}\sim (I_{LR},I_{HR})}{\Vert\nabla_{\Tilde{x}}(D(\Tilde{x})) - 1 \Vert}^2,
\end{split}
\end{equation}
where $\theta_D$ represents the trainable parameters of the discriminator.

\subsection {Two-Level Progressive Network}
Different from the neural networks that achieve SR imaging through one step, the progressive network splits an SR work into several sequential steps. In each step, the progressive network implements the image up-sampling by a small factor \cite{lai2017deep,Anwar2019}. Through the combination of several steps, it can implement the SR for a large up-sampling factor. LapSRN \cite{lai2017deep} is a good example of the progressive network, which contains three sequentially displaced subnets and each subnet can increase the image resolution by a factor of 2. Through such a neural network, LapSRN can finally improve image resolution by a factor of 8. 

Here we propose a two-level progressive neural network to achieve MCSR for an up-sampling factor of 4. The proposed model is the combination of two one-level nonprogressive network. Its generator contains two encoder-decoder networks and a reference feature extraction network, as shown in Fig. 1b. The discriminator is the same as the discriminator for the encoder-decoder network. To obtain 4-fold MCSR results, we trained the progressive network and ensured that it can progressively improve image quality level by level. In the first level, 2-fold MSCR results will be achieved. The other 2-fold MCSR results will be obtained in the second level. The input of the progressive network is the LR images through 4-fold down-sampling and zero-filling. The ground truth of the first level is the LR images through 2-fold down-sampling and zero-filling. The ground truth of the second level is the original HR images.

The objective function of the progressive network combines results of both the two levels. The modified equations of (4-6) are expressed as

\begin{align}
    \mathcal{L}_{mse} = & \mathbb{E}_{x\sim I_{4\times LR},y\sim I_{HR}}{\Vert G(G(x)) - y\Vert}^2\notag \\
    & + \mathbb{E}_{x\sim I_{4\times LR},y\sim I_{2\times LR}}{\Vert G(x) - y\Vert}^2\label{equ_9}\\ 
    \mathcal{L}_{per} = & \mathbb{E}_{x\sim I_{4\times LR},y\sim I_{HR}} \| \phi( G(G(x))) - \phi(y)\|^2 \notag \\
    & + \mathbb{E}_{x\sim I_{4\times LR},y\sim I_{2\times LR}}\| \phi(G(x)) - \phi(y)\|^2\label{equ_10}\\
    \mathcal{L}_{txt} = & \mathbb{E}_{x\sim I_{4\times LR},y\sim I_{HR}} \| GM[G(G(x))] - GM(y)\|^2\notag\\
    & + \mathbb{E}_{x\sim I_{4\times LR},y\sim I_{2\times LR}} \| GM(G(x)) \notag\\
    & - GM(y)\|^2,\label{equ_11}
\end{align}
where $I_{4\times LR}$ stands for 4-fold down-sampled LR images, $I_{2\times LR}$ for 2-fold down-sampled LR images, and $I_{HR}$ for original HR images.

\subsection {Image Quality Evaluation Metrics}
The structural similarity (SSIM), peak signal-to-noise ratio (PSNR) \cite{wang2004image} and information fidelity criterion (IFC) \cite{sheikh2005information} metrics are used to evaluate the image quality of MCSR results. 

SSIM is a fidelity metric that compares images in the terms of luminance, contrast, and structural similarity. A widely used SSIM expression is as follows:
\begin{equation}\label{equ_12}
    SSIM(x,y) = \frac{(2\mu_x\mu_y + c_1)(2\sigma_{xy} + c_2)}{(\mu_x^2 + \mu_y^2 + c_1)(\sigma_x^2 + \sigma_y^2 + c_2)},
\end{equation}
where $\mu_x$ and $\mu_y$ stand for the average pixel intensities of image $x$ and $y$, $\sigma_x^2$ and $\sigma_y^2$ represent the pixel intensity variances of $x$ and $y$. $\sigma_{xy}$ is the covariance of image $x$ and $y$, and $c_1$ and $c_2$ are two variables to stabilize the division. As SSIM measures the similarity at the patch level, in this work all SSIM scores will be given as the mean of SSIM values based on $7 \times 7$ patches.

PSNR is another commonly used fidelity measure. It is related to MSE, and can be expressed as the ratio between the square of the maximum value $MAX_I^2$ and the MSE value:

\begin{equation}\label{equ_13}
    PSNR(x,y) = 10\log_{10}\frac{MAX_I^2}{MSE}.
\end{equation}

IFC is proposed by Sheikh \cite{sheikh2005information}. It measures image quality based on natural scene statistics. While image information can be extracted by human visual system, IFC utilizes the mutual information between the reference and a distorted image to evaluate image quality. Compared with SSIM and PSNR, IFC takes the correlation between image and human visual perception into account. However, IFC is not the best at evaluating structural similarity.

Currently, there is no widely accepted measuring metric that can be used to accurately evaluated image quality in all possible circumstances. Each of the above-mentioned metric has its own limitation. To comprehensively evaluated image quality, these three metrics are used in combination.

\begin{figure*}[!ht]
\centering
\includegraphics[width=7.16in]{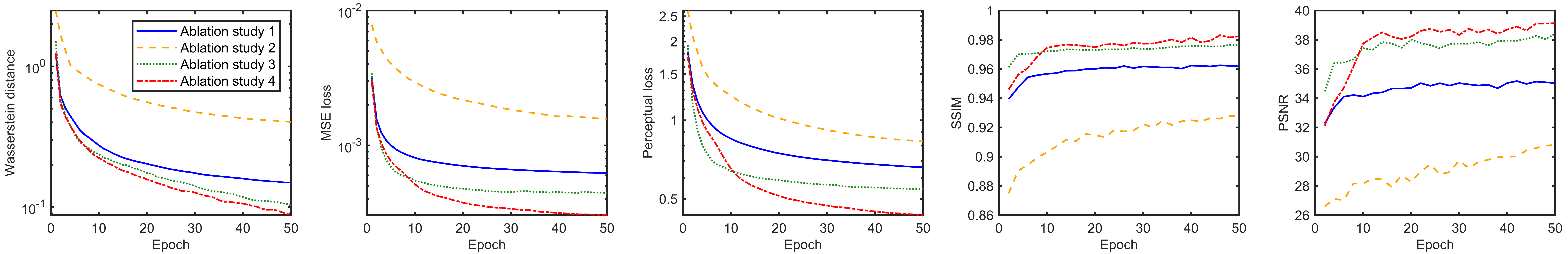}
\caption{Comparisons between different models used in our ablation studies in terms of the perceptual loss, mean-squared-error (MSE) loss, Wasserstein distance, PSNR and SSIM.}
\label{fig_2}
\end{figure*}

\section{Experiment}
\subsection {Datasets}
\textbf{IXI dataset:} The IXI dataset \cite{IXI} contains registered T2 weighted and PD weighted MRI images of 578 patients. T2 weighted images were used for SR with PD weighted images as the reference. 7,000 pairs of T2 and PD weighted images were selected for training. Another 1,955 pairs for testing. We adopted 10-fold cross-validation in this study. The size of original HR images of both T2 and PD weighted images is $256\times256$. 2-fold, 3-fold, and 4-fold down-sampled T2 weighted LR images were created through down-sampling and zero-filling. Before training, all images were normalized over the range of [0, 1]. Each LR and HR image pairs was cropped into 16 patch pairs with the patch size of $64\times64$. 

\textbf{NAMIC Dataset:} NAMIC brain multimodality dataset \cite{NAMIC} includes registered T1 and T2 weighted MRI images scanned from 20 patients.1,620 pairs were used for training, and 180 different pairs for testing. Similar to what we did with the IXI dataset, we chose T2 weighted images for SR with T1 weighted images as the reference. The size of the original HR image is $256\times256$. We down-sampled and zero-filled T2 weighted images for 2-, 3-, and 4-fold LR images. Before training, all images were normalized over the interval [0, 1]. Each LR and HR image pairs was cropped into 16 patch pairs with the patch size of $64\times64$.

\subsection {Experimental Details}
For training, 112K image patches of size $64\times64$ were randomly selected with a batch size of 32. The training process continued for 50 epochs with the learning rate of $2\times10^{-5}$. The discriminator was trained four times before the generator was trained once. Hyperparameters $\lambda_1$, $\lambda_2$, and $\lambda_3$ in the objective function of the generator were determined based on the result shown in Fig. 3. We first determined the coefficient of the perceptual loss. As shown in Fig. 3, both SSIM and PSNR scores get dramatically increased when $\lambda_2$ goes up from 0.01 to 1.0. Some further SSIM and PSNR improvement can be obtained by increasing $\lambda_2$ from 1.0 to 100. However, such an increment was far less than the increment in the range of 0.01 and 1.0. Considering a too large perceptual loss coefficient will overwhelm the contributions from other loss components, we finally set $\lambda_2$ to 1.0. Next, we tuned $\lambda_1$ and $\lambda_3$. For the coefficient of the MSE loss, when $\lambda_1$ was set to 0.1, the highest PSNR and the second highest SSIM scores can be obtained. Based on these, we set $\lambda_1$ to 0.1. Similarly, we set $\lambda_3$ to 0.1. The value of $\lambda_4$ in the objective function of the discriminator was 10, being consistent to the paper \cite{gulrajani2017improved}. The Adam optimization method \cite{kingma2014adam} was used to train the neural network. All the work was implemented using PyTorch on a GTX 1080Ti GPU. 

\begin{figure}[!h]
\centering
\includegraphics[width=3.5in]{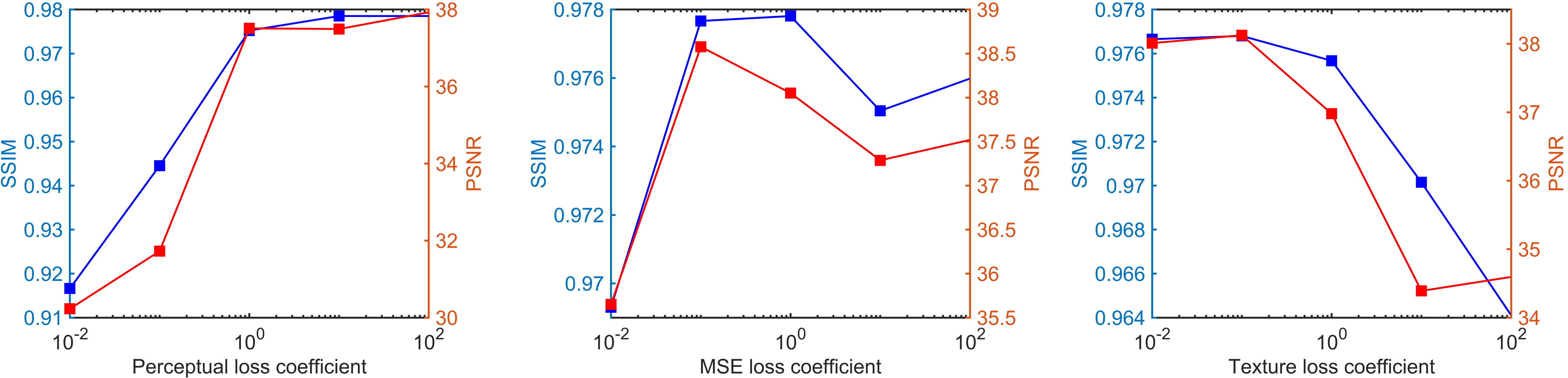}
\caption{Results of tuning hyperparameters in the objective function (the \textit{x}-axes are on logarithmic scale).}
\label{fig_3}
\end{figure}

\section{Results}
\subsection {Convergence Behavior}
During the training process, the perceptual loss $\mathcal{L}_{per}$, MSE loss $\mathcal{L}_{mse}$, and Wasserstein distance ($W_{dis}$) in each epoch were recorded for monitoring the convergence of the network model. Variations of the perceptual loss and MSE loss reflect the similarity between the generated result and the ground truth in the high-level feature space and low-level image space respectively. The Wasserstein distance is known as the earth-mover’s (EM) distance, which shows the probability distribution similarity between the generator output data and the true data. It is defined as 
\begin{equation}\label{equ_14}
    W_{dis} = \Big|\mathbb{E}_{x\sim I{LR}}[D(G(x))]-\mathbb{E}_{\hat{x}\sim I{HR}}[D(\hat{x})] \Big|.
\end{equation}

As shown in Fig. \ref{fig_2}, all perceptual loss curves, MSE loss curves, and Wasserstein distance curves gradually decrease towards zero with the continuation of the training process, indicating that our model converged stably. After 40 epochs, all curves become stable and close to zero, which reflects the fact that the generator can stably bring about super-resolution results with great similarity relative to the corresponding HR ground truth.

\begin{figure}[!h]
\centering
\includegraphics[width=3.5in]{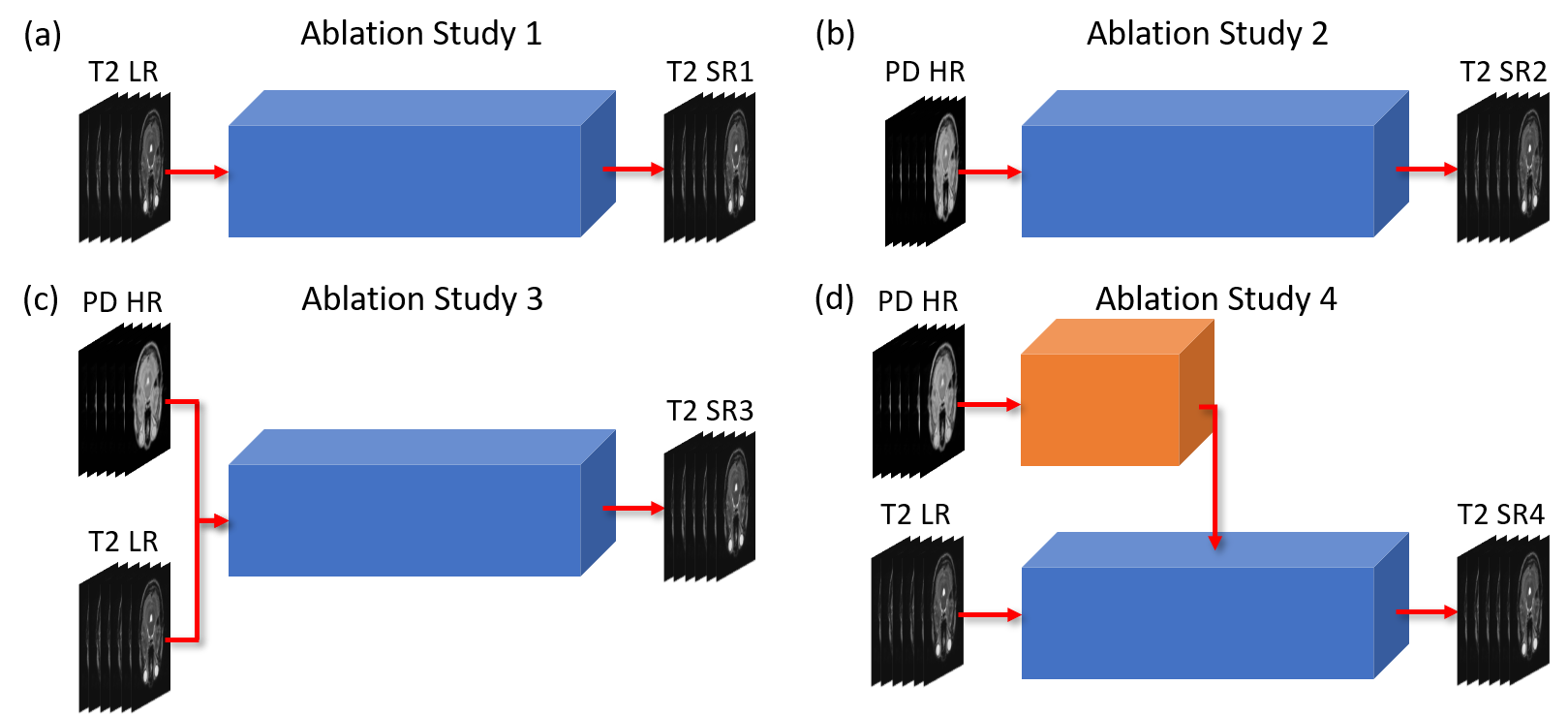}
\caption{Schematic views of the four ablation studies. (a)-(d) corresponds to the proposed four ablation studies. The blue block represents the encoder-decoder network and the orange block stands for the reference feature extraction network. (a) The first study is an SISR study in which the input is a T2 weighted LR image, and the output is an T2 weighted SR image. (b) The second study is an image synthesis study in which the input is a PD weighted HR image, and the output is a T2 weighted SR image. (c) The third study is an MCSR study in which the input is a T2 weighted LR image, with the PD weighted HR image as the reference, the output is the T2 weighted SR image. Multi-contrast information is combined in a low-level image space. (d) The fourth study is also an MCSR study in which the input is a T2 weighted LR image, with the PD weighted HR image as the reference, the output is the T2 weighted SR image. Multi-contrast information is combined in a high-level feature space.}
\label{fig_4}
\end{figure}

\subsection {Ablation Study}
\begin{figure}[!h]
\centering
\includegraphics[width=3.5in]{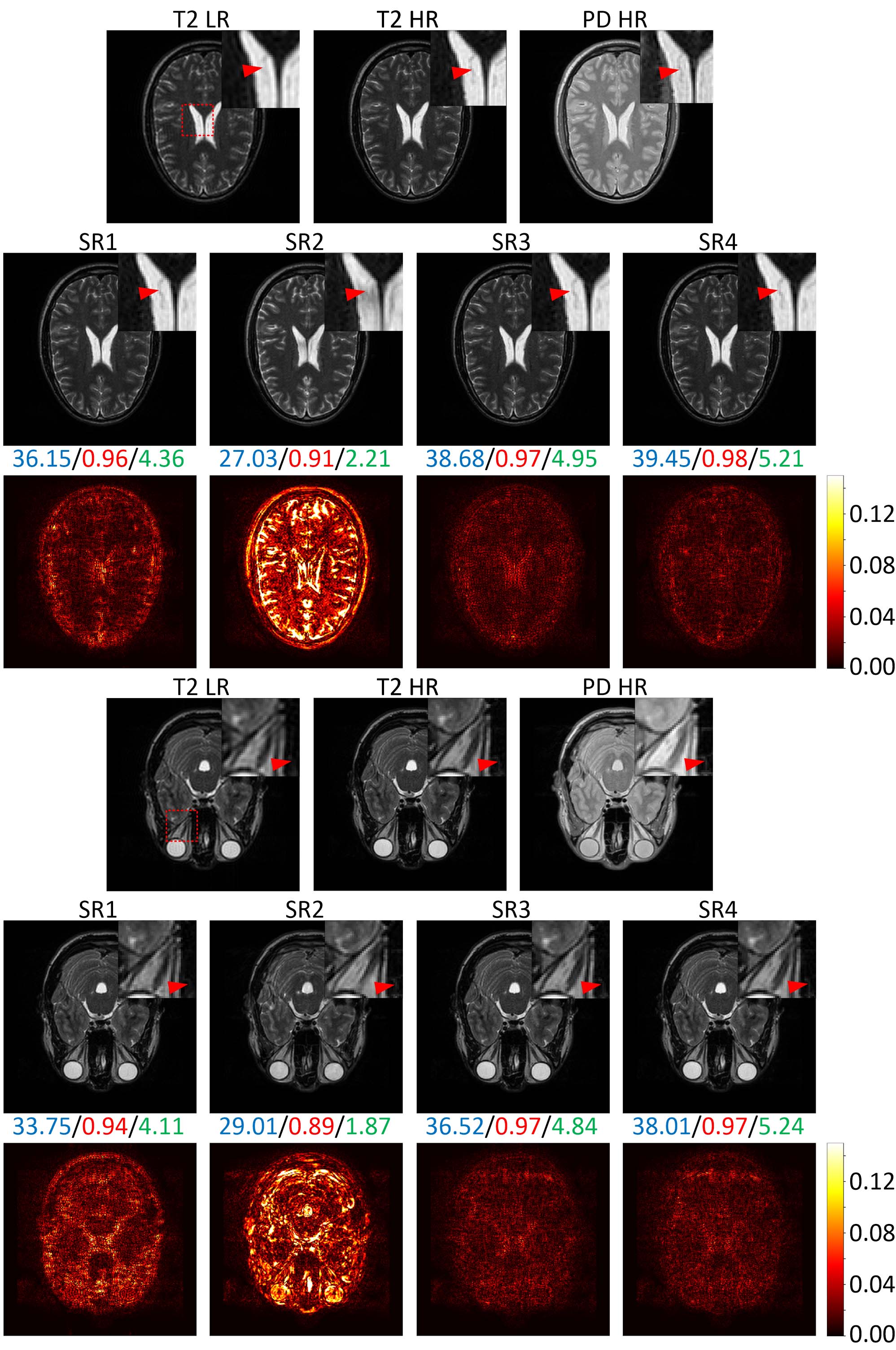}
\caption{Results from the ablation-based evaluation. PSNR, SSIM, IFC values below each SR result are shown in blue, red, and green, respectively. In the second and fifth row, SR1, SR2, SR3, and SR4 indicate the results from four ablation studies respectively. The hot maps in the bottom row show the absolute pixel-value difference between ablation study results and the corresponding ground truth T2 weighted HR image.}
\label{fig_5}
\end{figure}

To find the best neural network structure for MCSR imaging, we built four models and conducted a systematic ablation evaluation. In this subsection, only IXI dataset was used. In the first study, to investigate the utility of the introduction of reference images, we removed the reference feature extraction network from the proposed network and only kept a single encoder-decoder network with only T2 weighted LR images as the input. In the second study, we were interested in directly converting PD weighted images into T2 weighted images through an image synthesis process. Hence, we used PD weighted HR images as the input to the same neural network in the first study, instead of T2 weighted LR images. In the third and fourth studies, we focused on how to effectively utilize the information in PD weighted HR images. In the third study, these two-contrast images were combined in a low-level image space and fed into the same neural network model used in the previous two studies. In the fourth study, two-contrast images were combined in a high-level feature space, which is the proposed network. The above four studies are illustrated in Fig. \ref{fig_4}. For comparison, all input T2 weighted LR images used in the ablation study are 2-fold down-sampled LR images. Results from the ablation studies are shown in Fig. \ref{fig_5} and Table \ref{table_1}. 

Fig. 2 shows that convergence curves in the second ablation study, with the highest Wasserstein distance, perceptual loss, MSE loss and the lowest SSIM and PSNR values. As shown in Fig. 5, the SR result obtained in the second study also has the worst image quality as it has the largest pixel-value difference from the ground truth. Taking the center white matter regions in the first three rows as an example, all other SR results show the fissure with clear shapes while the second SR result cannot. Quantitative results in Table \ref{table_1} also indicate that the results from the second ablation study have the lowest SSIM, PSNR and IFC scores. All these results emphasize that the image synthesis study converting PD weighted HR images into T2 weighted HR images cannot obtain high-quality SR results. 

When comparing results among the first, third and fourth ablation studies, MCSR results from the third and fourth ablation study beat the SISR results from the first study. MCSR results have higher PSNR, SSIM, and IFC scores and less pixel-value differences between MCSR results and the ground truths than SISR results. As an example, the dark curve highlighted by the red arrow in the T2 weighted LR image in the fourth row of Fig. \ref{fig_5} can be hardly distinguished from the background. However, this curve shows clearly in the corresponding PD weighted HR image. After the training, this curve is still hard to be seen in the SR result in the first ablation study but it became much clearer in the third and fourth ablation studies that utilized PD weighted HR images in the training process. 

In the third and fourth ablation study, curves of the third study in Fig. \ref{fig_2} decrease quickly in the first several training epochs. However, after 10 epochs, curves of the fourth study are slightly lower than the third study. Fig. \ref{fig_5} demonstrates that the fourth study can achieve SR results with smaller pixel-value differences. Quantitative results in Table I also show that the fourth ablation study results beat the third ablation counterparts in terms of PSNR, SSIM and IFC scores.

\begin{table}[!h]
\caption{Statistic Result of Ablation Studies Based on IXI Dataset (MEAN$\pm$STD)}
\label{table_1}
\centering
\begin{tabular*}{\columnwidth}{  @{\extracolsep{\fill}} c c c c c @{} }
\toprule
{} & SR1 & SR2 & SR3 & SR4\\  
\midrule
SSIM  & 0.960$\pm$0.012 & 0.915$\pm$0.025 & 0.975$\pm$0.006 & 0.976$\pm$0.006 \\
PSNR & 34.705$\pm$1.457 & 28.282$\pm$1.544 & 38.225$\pm$1.684 & 38.513$\pm$1.765 \\ 
IFC & 3.990$\pm$0.246 & 1.890$\pm$0.178 & 4.666$\pm$0.298 & 5.058$\pm$0.332 \\
\bottomrule
\end{tabular*}
\end{table}

\subsection {Non-progressive Model MCSR Results with 2-fold, 3-fold and 4-fold Up-sampling}
\begin{figure}[!h]
\centering
\includegraphics[width=3.5in]{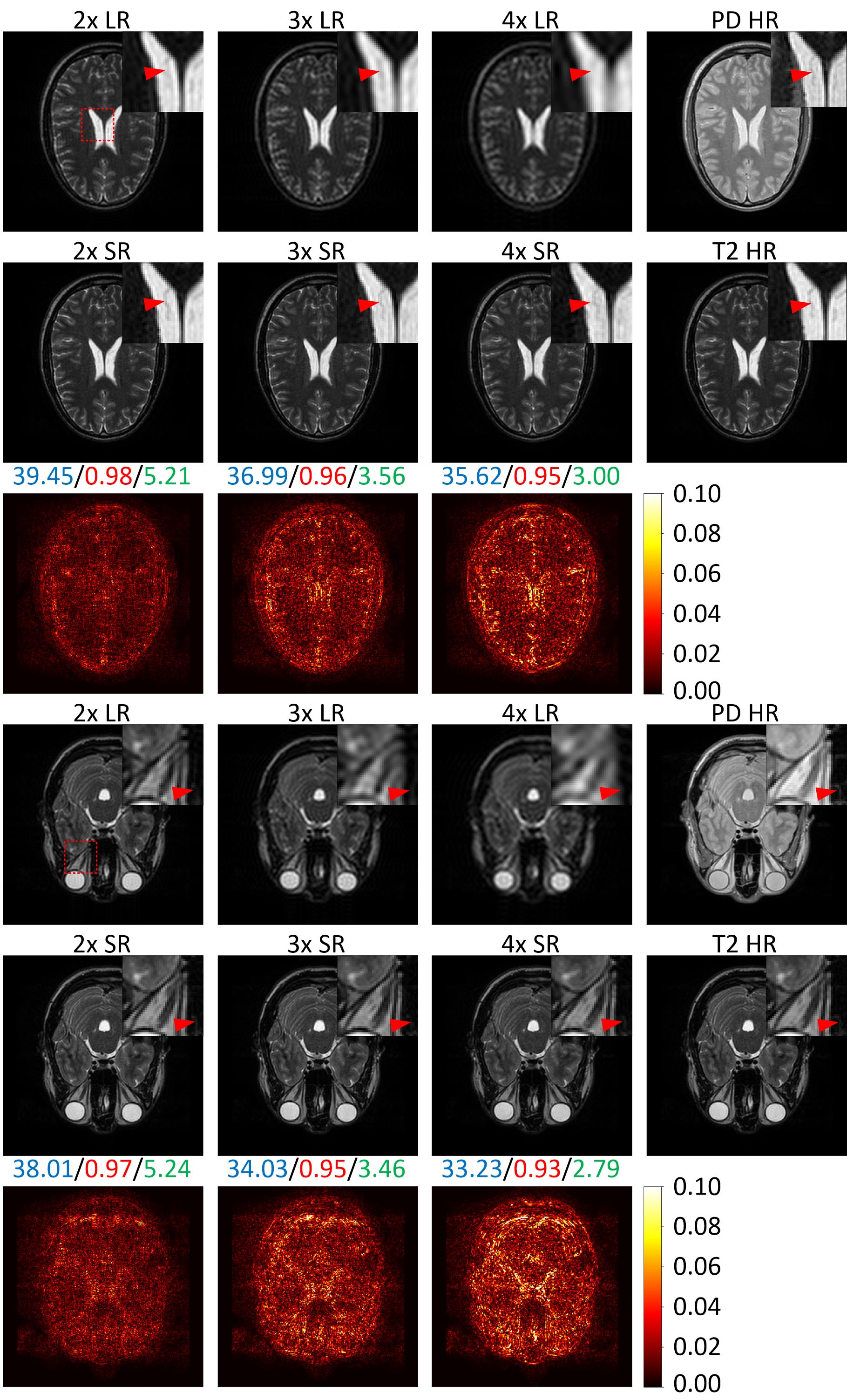}
\caption{T2 weighted SR results with different down-sampling factors based on the IXI dataset. LR images with different down-sampling factors are shown in the first and fourth rows, and their corresponding SR results in the second and fifth rows. PSNR, SSIM, and IFC scores are in blue, red, and green respectively below each subfigure. The hot maps show the absolute pixel-value differences between MCSR results and the ground truth T2 weighted HR image.}
\label{fig_6}
\end{figure}

\begin{figure}[!h]
\centering
\includegraphics[width=3.5in]{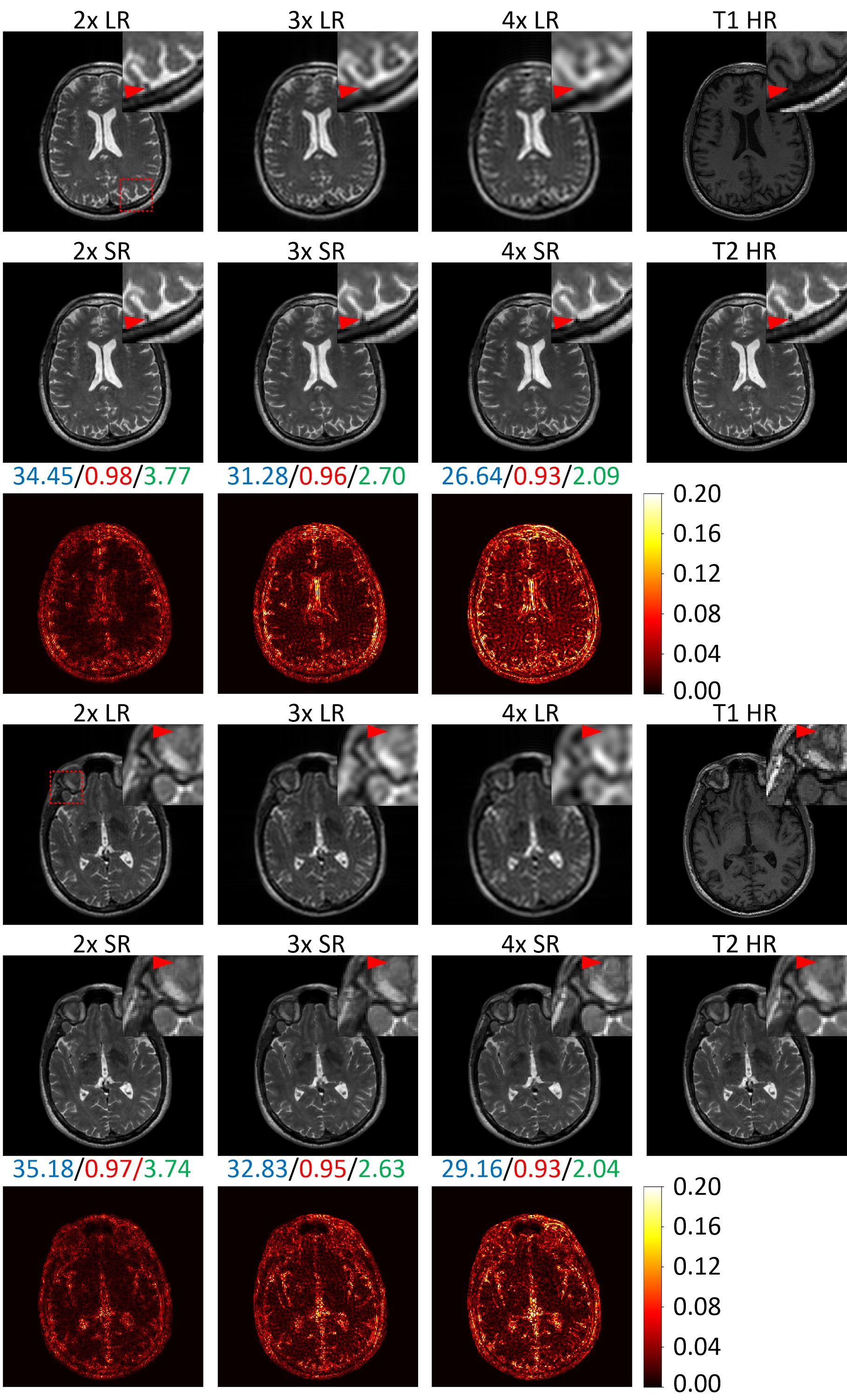}
\caption{T2 weighted SR results with different down-sampling factors based on the NAMIC dataset. LR images with different down-sampling factors are shown in the first and fourth rows and their corresponding SR results in the second and fifth rows. PSNR, SSIM, and IFC scores are in blue, red, and green respectively below each subfigure. The hot maps show the absolute pixel-value differences between the corresponding MCSR results and the ground truth T1 weighted HR images.}
\label{fig_7}
\end{figure}

In the previous subsection, we demonstrated that our method can achieve 2-fold up-sampling MCSR results with good image quality. Now, let us study MCSR with a higher up-sampling factor. MCSR results for 2-, 3-, and 4-fold up-sampling rates based on the one-level model are shown in Figs. \ref{fig_6}, \ref{fig_7}, and Table \ref{table_2}. It can be seen that when images are down-sampled by a larger factor, the down-sampled LR images become much more blurred with much less visible details. After training, we can obtain quite good MCSR results with clearer shapes and textual details, and these MCSR results are with great structural similarity with the corresponding HR image. For example, fissures in the white matter pointed by red arrows in Fig. 6 become gradually less clear and less detectable with the increment of down-sampling factors. However, in each MCSR result, these fissures become much clearer and detectable. Among all MCSR results, it can be found that MCSR results obtained from a smaller down-sampling factor have higher PSNR, SSIM, and IFC scores and less pixel-value differences with the HR ground truth, indicating that these MCSR results based on LR images with a small down-sampling factor have higher image qualities than those based on LR images with a larger down-sampling factor, as heuristically expected.

\begin{table}[!h]
\caption{Statistic Result of Larger-Factor MCSR Imaging Using The One-Level Non-progressive Model (MEAN$\pm$STD)}
\label{table_2}
\centering
\begin{tabular*}{\columnwidth}{  @{\extracolsep{\fill}} c c c c c @{} }
\toprule
{} & {} & 2$\times$SR & 3$\times$SR & 4$\times$SR \\  
\midrule
\multirow{3}{0.8cm}{\centering IXI Dataset} & SSIM  & 0.976$\pm$0.006 & 0.960$\pm$0.010 & 0.950$\pm$0.013 \\
                             & PSNR & 38.513$\pm$1.765 & 35.215$\pm$1.820 & 33.329$\pm$2.116 \\
                             & IFC & 5.058$\pm$0.332 & 3.304$\pm$0.245 & 2.708$\pm$0.215 \\
\hline
\multirow{3}{0.8cm}{\centering NAMIC Dataset} & SSIM  & 0.974$\pm$0.006 & 0.954$\pm$0.010 & 0.925$\pm$0.014 \\
                             & PSNR & 34.127$\pm$0.978 & 31.562$\pm$1.149 & 28.207$\pm$1.178 \\
                             & IFC & 3.538$\pm$0.164 & 2.504$\pm$0.105 & 1.923$\pm$0.113 \\
\bottomrule
\end{tabular*}
\end{table}

\subsection {Progressive Model Further Improves Image Quality with 4-fold Up-sampling}
In the previous subsection, although quite good MCSR results can be obtained from highly down-sampled LR images like 4-fold SR results, these results are still with lower image quality compared with results based on low down-sampling factor LR images. We hope to further improve the quality of highly down-sampled LR images, and proposed a two-level progressive model to achieve this goal. In this subsection, we presented two MCSR results from the proposed progressive model trained with different objective functions. One is called 4$\times$PRO U SR, and the other one 4$\times$PRO C SR. ``U'' and ``C'' mean unconstrained and constrained respectively. The unconstrained results are from the progressive model trained with the objective functions \eqref{equ_4}, \eqref{equ_5}, and \eqref{equ_6} that only considers the final 4-fold SR results. On the other hand, the constrained results are from the progressive model that trained with the objective functions \eqref{equ_9}, \eqref{equ_10}, and \eqref{equ_11} that considers both 2-fold and 4-fold SR results. 

\begin{figure}[!h]
\centering
\includegraphics[width=3.5in]{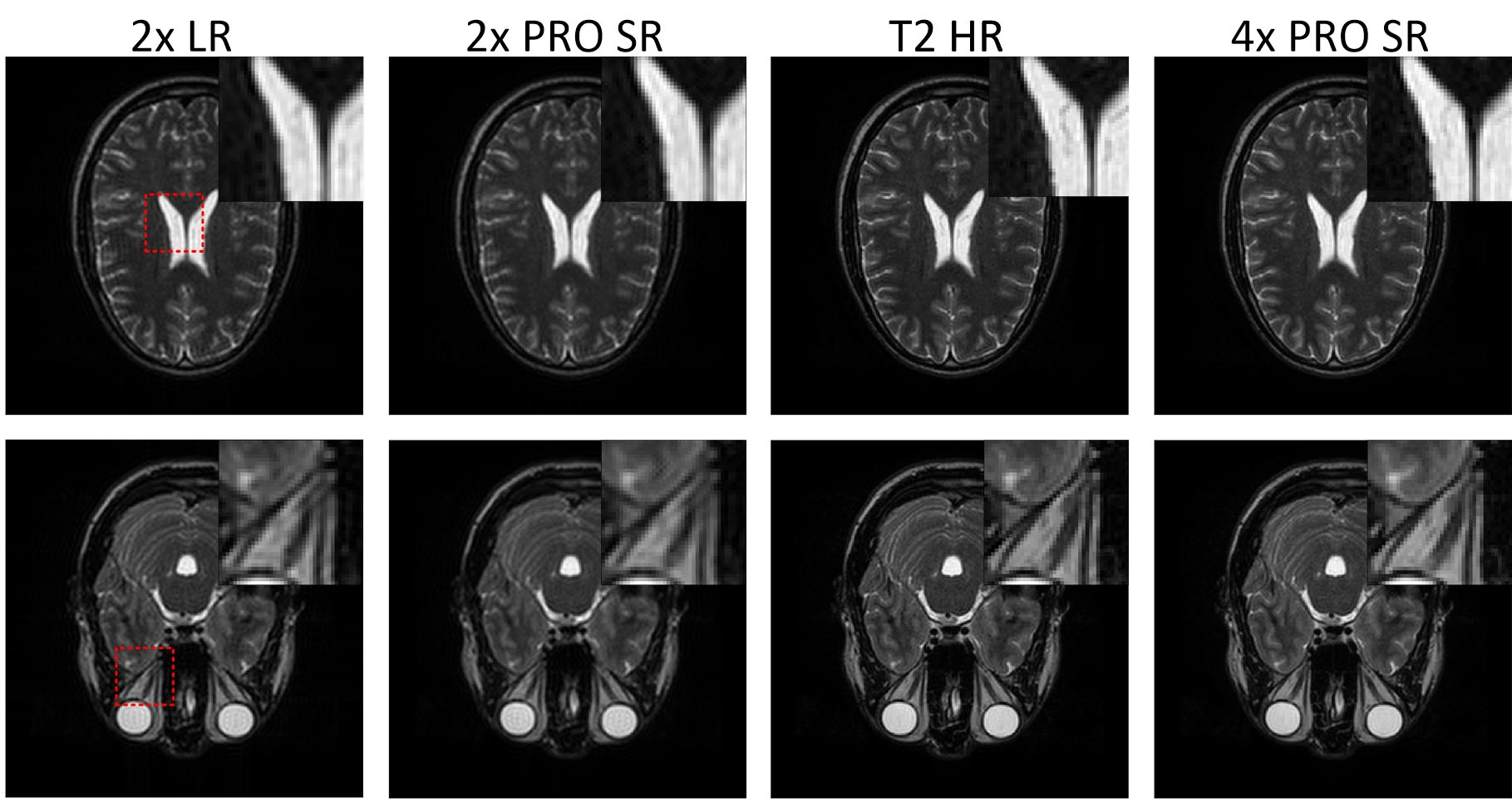}
\caption{Comparison of the two-level progressive network outputs with their ground truths. 2$\times$LR represents the ground truth at the first level, and T2 HR stands for the ground truth at the second level. 2$\times$PRO SR shows the outputs of the first level, and 4$\times$PRO SR the outputs of the second level.}
\label{fig_8}
\end{figure}

Fig. \ref{fig_8} shows the comparison between each subnet’s output and its corresponding ground truth based on the 4$\times$PRO C SR model. As shown in Fig. \ref{fig_8}, our two-level progressive network can generate two levels of MCSR results under a strong restriction. Compared with their ground truths, the network can produce very faithful results. 

\begin{figure}[!h]
\centering
\includegraphics[width=3.5in]{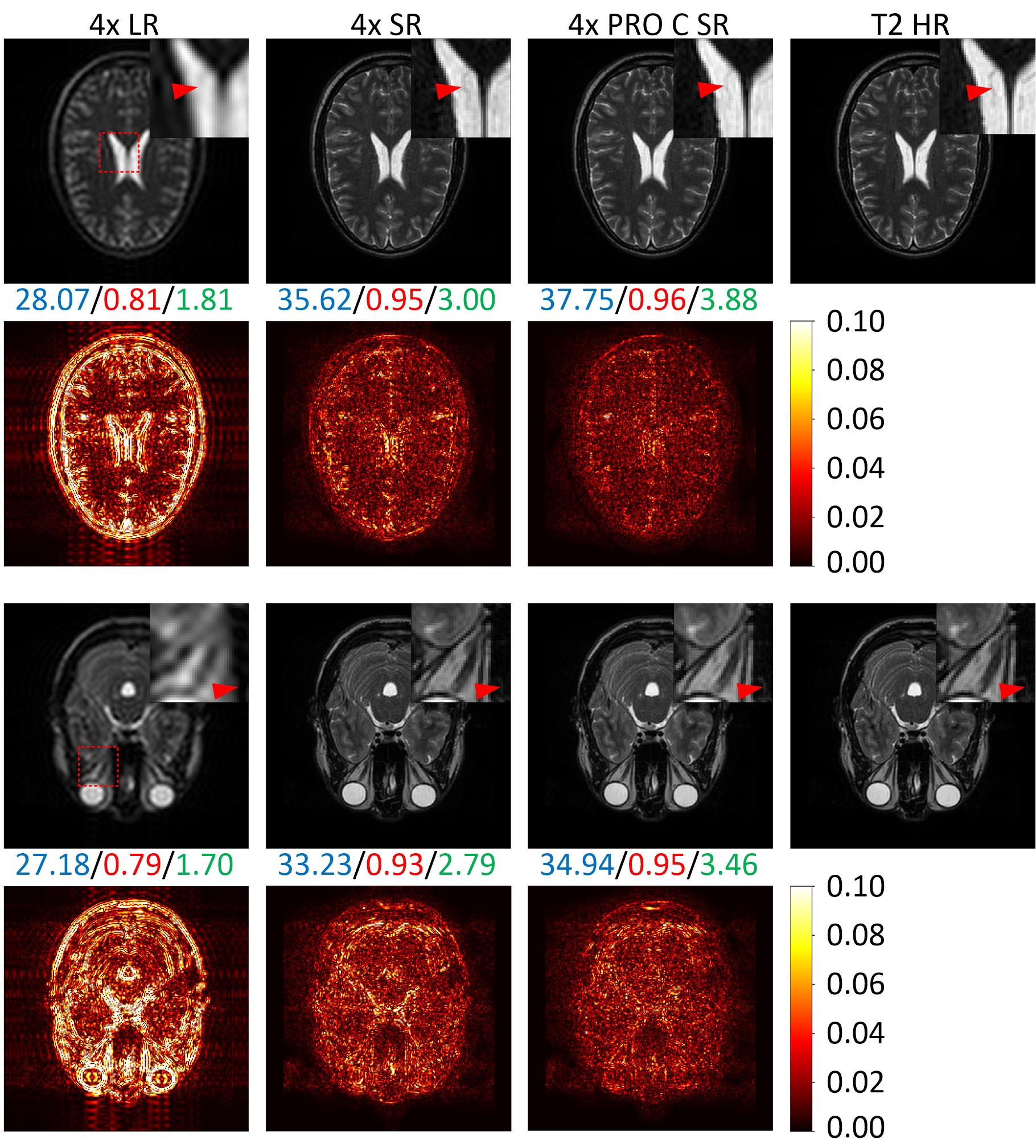}
\caption{Comparisons of T2 weighted MCSR results from the one-level non-progressive network and the two-level progressive network. Results are all based on the IXI dataset. 4$\times$SR and 4$\times$PRO C SR denote MCSR results using the one-level non-progressive network and the two-level progressive network respectively. PSNR, SSIM, and IFC scores are in blue, red, and green respectively below each subplot. The hot maps show the absolute pixel-value differences between the super-resolution results and the ground truth T2 weighted HR images.}
\label{fig_9}
\end{figure}

\begin{figure}[!h]
\centering
\includegraphics[width=3.5in]{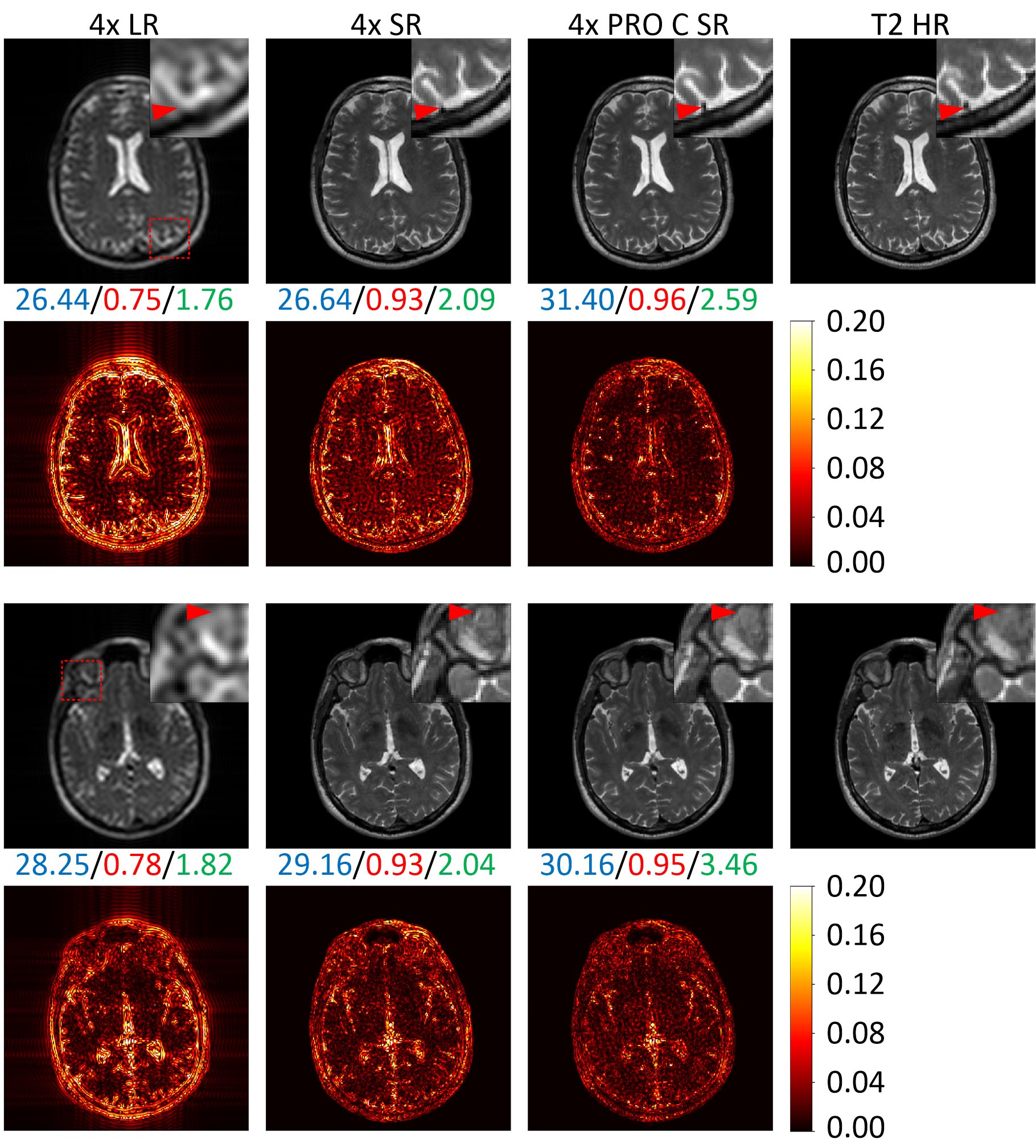}
\caption{Comparisons of T2 weighted MCSR results from the one-level non-progressive network and the two-level progressive network. Results are all based on the NAMIC dataset. 4$\times$SR and 4$\times$PRO C SR denote MCSR results using the one-level non-progressive network and the two-level progressive network respectively. PSNR, SSIM, and IFC scores are in blue, red, and green respectively below each subfigure. The hot maps show the absolute pixel-value differences between corresponding results and the ground truth T1 weighted HR images.}
\label{fig_10}
\end{figure}

Figs. \ref{fig_9} and \ref{fig_10} show the comparison of 4-fold MCSR results based on the one-level non-progressive model and the two-level progressive model. It can be found that the progressive network produced MCSR results have sharper edges, clearer texture details, and are with higher PSNR, SSIM, and IFC scores and less pixel-value differences from the corresponding ground truths than the one-level non-progressive model. Table \ref{table_3} statistically compares different methods. Both the 4$\times$PRO U SR and 4$\times$PRO C SR progressive results are better than the non-progressive results. Compared with the progressive result from the unconstrained model (4$\times$PRO U SR), the result from the constrained model (4$\times$PRO C SR) have higher SSIM, PSNR, and IFC scores, indicating the constrained progressive model can bring about SR imaging result with better image quality than the unconstrained model.

\begin{table*}[!t]
\caption{Statistical Comparison of MCSR Results From Non-progressive Model And Progressive Models (MEAN$\pm$STD)}
\label{table_3}
\centering
\begin{tabular}{>{\centering\arraybackslash}m{2cm}>{\centering\arraybackslash}m{2cm}>{\centering\arraybackslash}m{2.5cm}>{\centering\arraybackslash}m{2.5cm}>{\centering\arraybackslash}m{2.5cm}>{\centering\arraybackslash}m{2.5cm}}
\toprule
 &  & 4$\times$LR & 4$\times$SR & 4$\times$PRO U SR & 4$\times$PRO C SR \\  
\midrule
\multirow{3}{0.7cm}{\centering IXI Dataset} & SSIM  & 0.793$\pm$0.029 & 0.950$\pm$0.013 & 0.958$\pm$0.011 & 0.961$\pm$0.010 \\ 
                             & PSNR & 27.151$\pm$1.309 & 33.329$\pm$2.116 & 34.814$\pm$2.192 & 35.140$\pm$2.210 \\
                             & IFC & 1.627$\pm$0.150 & 2.708$\pm$0.215 & 3.129$\pm$0.270 & 3.354$\pm$0.303 \\
\hline
\multirow{3}{0.7cm}{\centering NAMIC Dataset} & SSIM  & 0.751$\pm$0.024 & 0.925$\pm$0.014 & 0.951$\pm$0.011 & 0.953$\pm$0.011 \\
                             & PSNR & 26.827$\pm$1.131 & 28.207$\pm$1.178 & 30.763$\pm$1.229 & 30.942$\pm$1.584 \\
                             & IFC & 1.626$\pm$0.120 & 1.923$\pm$0.113 & 2.352$\pm$0.118 & 2.385$\pm$0.120 \\
\bottomrule
\end{tabular}
\end{table*}

\subsection {Comparing with State-of-the-art Methods}
We also compared our results with some recently published state-of-the-art (SOTA) MCSR methods, including SSIP \cite{manjon2010mri}, SRGR \cite{zheng2018multi}, and Zeng’s model \cite{zeng2018simultaneous}. According to the results shown in Table \ref{table_4}, our one-level-nonprogressive model achieved the highest SSIM scores among all the methods. the progressive model produced the highest SSIM and PSNR scores among all the methods.

\begin{table}[!h]
\caption{Comparisons of Our Methods with Other SOTA MCSR Methods Based on The NAMIC Dataset}
\label{table_4}
\centering
\begin{threeparttable}
\begin{tabular}{>{\centering\arraybackslash}m{0.9cm}>{\centering\arraybackslash}m{0.9cm}>{\centering\arraybackslash}m{0.9cm}>{\centering\arraybackslash}m{1.2cm}>{\centering\arraybackslash}m{1.2cm}>{\centering\arraybackslash}m{1.2cm}}
\toprule
UP-sampling Factor & SSIP* & Zeng* & SRGR** & Non-progressive Model & Progressive Model \\  
\midrule
\multirow{2}{*}{$\times$2} & 0.924 & 0.945 & 0.959 & \textbf{0.974} & - \\
                   & 25.34 & \textbf{38.32} & 33.98 & 34.13 & - \\
\hline
\multirow{2}{*}{$\times$3} & 0.790 & 0.872 & - & \textbf{0.954} & - \\
                   & 21.08 & \textbf{33.76} & - & 31.56 & - \\
\hline
\multirow{2}{*}{$\times$4} & 0.663 & 0.811 & - & 0.925 & \textbf{0.953} \\
                   & 18.90 & 30.84 & - & 28.20 & \textbf{30.94} \\
\bottomrule
\end{tabular}
\begin{tablenotes}
\item[] In each line, top values are SSIM scores and bottom values are PSNR scores.
\item[*] Results are from Table II in \cite{zeng2018simultaneous}.
\item[**] Results are based on the averaging of two MCSR results listed in Table I in \cite{zheng2018multi}.
\end{tablenotes}
\end{threeparttable}
\end{table}

\section{Discussions}
In this study, we have jointly used the adversarial loss, MSE loss, perceptual loss, and textual matching loss in the objective function. Table \ref{table_5} has summarized the comparison of SR imaging results with different objective functions. The results obtained with all four loss components enjoy the highest SSIM score and the second highest PSNR score. Adding the textual matching loss decreases the PSNR score, which is in accordance with the results in \cite{sajjadi2017enhancenet}. This can be explained by the fact that the MSE loss directly contributes to the PSNR score. Adding the texture matching loss dilates the contribution of the MSE loss term. Despite adding the texture matching loss lowers the PSNR score, SR imaging results actually are with better overall image quality. Here, we support this point by showing that the highest SSIM score can be obtained by using all four loss functions together, which is in favor of the results presented in the paper \cite{sajjadi2017enhancenet}.

\begin{table}[!h]
\caption{Comparisons of SR Results on IXI Dataset Through Using Different Combinations of Loss Functions in the Objective Function}
\label{table_5}
\centering
\begin{tabular}{>{\centering\arraybackslash}m{4.5cm}>{\centering\arraybackslash}m{1.5cm}>{\centering\arraybackslash}m{1.5cm}}
\toprule
 &  SSIM & PSNR \\  
\midrule
$\mathcal{L}_{adv}+\mathcal{L}_{per}$ & 0.973 & 37.899 \\
$\mathcal{L}_{adv}+\mathcal{L}_{per}+\mathcal{L}_{mse}$ & 0.975 & 38.966 \\
$\mathcal{L}_{adv}+\mathcal{L}_{per}+\mathcal{L}_{mse}+\mathcal{L}_{txt}$ & 0.976 & 38.513 \\
\bottomrule
\end{tabular}
\end{table}

To investigate if the combination of different contrast image will help to improve the SR result and which information combination can achieve the best result, we conducted four ablation studies. According to the ablation results shown in Figs. \ref{fig_3}, \ref{fig_5} and Table I, the use of only PD weighted HR images for image synthesis cannot generate good T2 weighted images, which indicates that PD weighted HR images cannot provide enough information or image priors for the neural network to reconstruct a T2 weighted image of high quality. Compared with the SISR method, MCSR methods show superior results, suggesting that multi-contrast images can provide much more image priors than single-contrast images for super-resolution imaging.

When comparing results of the third ablation study to that of the fourth study, the results from the fourth study slightly outperform that from the third one. With respect to different multi-contrast information combinations, the resultant SR imaging results indicate that multi-contrast image information combined in a high-level feature space can bring about better MCSR results than a low-level image space combination. These results may be appreciated from three aspects. First, adding such a reference feature extraction network splits the feature extraction process of two-contrast images, which facilities the neural network to flexibly extract features based on each contrast modality’s unique characteristics. Second, more parameters are used in the high-level combination strategy, which indicates that the neural network may have a stronger ability to extract more multi-contrast features for the SR imaging. Third, the high-level combination model only transfers features of T2 weighted images through skip connections to the decoder. Such a design enhances the contribution of T2 weighted features in the image restoration process happened in the decoder, and makes sure that the SR images are restored with more T2 characteristics.  

Results of the one-level nonprogressive model in Fig. \ref{fig_6} and Table II show that our method produces MCSR results with clear edges and textures even if the original LR images are highly down-sampled with severe information loss. However, the image quality of MCSR results decreases with the increment of the down-sampling factor, which is not surprising as further down-sampled LR images come with less details, and it would be more difficult for the neural network to recover details based on such more blurred images. 

To obtain good SR results with high up-sampling factors, we propose a two-level progressive network to progressively achieve MCSR level-/stage-wise. Compared with 4-fold MCSR results obtained from a one-level nonprogressive network, the proposed two-level progressive network shows significantly better results. According to Table IV, the progressive model can even achieve 4-fold up-sampling MCSR results close to 3-fold up-sampling results from the non-progressive model. The improvement of the progressive model attributes to the strong model parameter restriction. Parameters in the generator of the progressive network are strictly constrained during the training process guided by the objective function. At each level, the encoder-decoder network is trained to produce 2-fold up-sampling SR results. Compared with a loosely constrained training process like the one that only makes sure the final output close to the ground truth and no constraint on intermediate results (4$\times$PRO U SR in Table III), such a strong constriction can boost the optimization process and ensure parameters more reasonably adapted to generate good results. 

Our progressive model can be extended to perform larger factor up-sampling MCSR tasks. For example, our current two-level progressive model can be extended into a three-level model for 8-fold up-sampling MCSR. We believe that for 8-fold resolution improvement, more contrast mechanisms may be needed. For example, we may need to couple CT and MRI images together \cite{wang2015vision}. We have not implemented such an 8-fold MCSR because the limitation of the dataset. Also, the size of original HR images used in this study is 256$\times$256, with 8-fold down-sampling the obtained LR images would be too blurry.

\section{Conclusion}
In this paper, we have proposed a one-level non-progressive neural network for low factor up-sampling (such as 2$\times$) and a two-level progressive neural network for large factor up-sampling (such as 4$\times$). For achieving better results, multiple losses have been combined in a composite object function. According to our results, MCSR results with high image quality can be generated when multi-contrast information is combined in a high-level feature space. The proposed progressive network could be extended for MCSR results with an even larger up-sampling factor if more different contrast images are available.

\ifCLASSOPTIONcaptionsoff
  \newpage
\fi

\bibliographystyle{IEEEtran}
\bibliography{IEEEabrv,main}

\vfill

\end{document}